\newcommand{\be}{\begin{equation}}
\newcommand{\ee}{\end{equation}}
\newcommand{\bea}{\begin{eqnarray}}
\newcommand{\eea}{\end{eqnarray}}

\newcommand{\bra}[1]{\big< #1 \big|}
\newcommand{\ket}[1]{\big| #1 \big>}

\newcommand{\nn}{\nonumber}
\documentstyle[11pt]{article}
\title{The connection between single transverse spin asymmetries
        and the second moment of $g_2$}
\author{B.~Ehrnsperger, A.~Sch\"afer, W.~Greiner
\\
Institut f\"ur Theoretische Physik, J.~W.~Goethe
Universit\"at Frankfurt,
\\
Postfach~11~19~32, W-60054~Frankfurt am Main, Germany
\\
and L.~Mankiewicz
\\
N. Copernicus Astronomical Center, Bartycka 18, PL--00--716, Warsaw, Poland
\\
UFTP preprint 348/1993     hep-ph/9312264 }
\begin{document}
\maketitle
\begin{abstract}
We point out that the size of the photon  single spin asymmetry
in high--energy proton
proton collisions with one transversely polarized proton can be related
to $d^{(2)}$, the twist
three contribution  to the second moment of $g_2$. Both quantities should
be measured
in the near future. The first was analysed by Qiu and Sterman, the second
was estimated by Balitsky,
Braun, and Kolesnichenko. Both experiments measure effectively the
strength of the
collective gluon field in the nucleon oriented relative to the nucleon spin.
The sum rule results
suggest that the single spin asymmetry is rather small for the proton,
but could be substantial for the neutron.
\end{abstract}
Recent polarized deep inelastic lepton nucleon scattering
experiments \cite{BAd93,PAn93,JAs89} have generated great interest in
the spin structure of the nucleons.
It seems that polarized reactions will develop into an ideal testing ground
for QCD.
So far only longitudinal polarization was studied, but transverse polarization
could offer a wealth of novel effects and will be studied in detail by future
experiments.
In deep inelastic scattering the interest is focused on the polarized
distribution function $g_2(x)$ which was already extensively analysed
theoretically \cite{RJa90,RJa91,LMa91,ASc92}.
In these investigations unsuppressed twist three
contributions $d^{(n)}$ to the $n$-th moment of $g_2(x)$
are most interesting.
$d^{(2)}$ should be measurable with good statistics e.g. by the
HERMES \cite{Hermes} experiment.
In priciple a plethora of transverse spin effects should be observable in
polarized proton-proton collisions.
However, the expected asymmetries in these reactions are usually very small,
typically of the order of 1 \% or less.
Furthermore it will still take a number of years before polarized
proton-proton collisions become feasible at any accelerator, the ideal
place definitely being RHIC.
What is possible in the near future are single spin experiments with a
polarized target and an unpolarized beam and these are
under intensive consideration.

In this situation it is most interesting that
Qiu and Sterman (QS) pointed out the possibility of a large single--spin
asymmetry (of the order of 10 -- 20 \%)  for high--transverse--momentum
direct--photon production in pp collisions \cite{JQi91}.
They consider the following process:
\be \label{eq1a}
N(p,s_T) + N'(p') \to \gamma (l) + X
\ee
$N(p,s_t)$ and $N'(p')$ represent a transversely polarized nucleon of
momentum $p$ ($p^2 = M^2$) and spin $s_T$ ($s_T^2 = -1$),
and an unpolarized nucleon of momentum
$p'$.
$l$ is the momentum of the photon, $E_l$ its energy,
$l_T$ the transverse component of $l$.
The fractional asymmetry $A$ under consideration is
\be
A(s_T , x_F, l_T ) = E_l \frac{d \Delta \sigma_\perp }
{d^3 l }
/
\frac{E_l d \sigma }
{d^3 l }
\; ,
\ee
with
\be
\Delta \sigma_\perp = \frac{1}{2}
\left( \sigma (s_T , l) - \sigma (- s_T , l) \right) \; ,
\ee
and $\sigma (s_T ,l)$ denotes the direct--photon--production cross section
for the scattering of Eq. (\ref{eq1a}).
As typical values for the reaction QS take $\sqrt{s} =
30 $ GeV, $l_T = 4$ GeV .
QS showed that the leading contribution to the process (\ref{eq1a})
comes from a twist 3 parton distribution $T(x,s_T)$,
involving the correlation between quark
fields and the gluonic field strength.
With our conventions ($D_\mu = \partial_\mu + {\rm i} g A_\mu $)
and in the $n \cdot A = 0$ gauge it
has the form \cite{GS93}:
\be \label{eq1}
T(x,s_T) = - \int \frac{dy_1^-}{4 \pi} e^{ixp^+ y_1^-}
           \bra{p,s_T} \bar{\psi}(0) \gamma^+
            \int dy_2^- \varepsilon_{\sigma \rho \alpha \beta}
                 S_T^\sigma n^\alpha \bar{n}^\beta G^{\rho +} (y_2^-)
           \psi (y_1^-) \ket{p,s_T}  \; .
\ee
$\bar{n}^\beta = \delta_{\beta +}$ and
$n^\alpha = \delta_{\alpha -}$.

In general gauge factors for the color parallel transport from
$y_1^-$ to $y_2^-$ and $y_2^-$ to 0 have to be inserted.
The asymmetry $A$ is directly proportional to the size of this
matrix element.

The main problem of most of the various imaginable hadronic spin
asymmetries is that of interpretation.
Let us only remind of the variety of single
spin asymmetry measurements \cite{GBu76},
the physical interpretation of which is still controversial.
The nice thing about the single--spin photon asymmetry is that it can be
linked to a clear physical origin, namely the possible existence of a
collective gluon field in the interior of the nucleon.
One part of this field, e.g. a color magnetic field, is parallel to the
nucleon spin.
For such a constant field no destructive interference occures.
This is incorporated in Eq. (\ref{eq1})
by setting the momentum associated with $y_2^-$ equal to zero.
As a collective field is such a clear physical feature one should expect
that it contributes to a variety of observables.
This is in fact true as we shall show now.

To proceed we first simplify Eq. (\ref{eq1}).
We choose the target rest system for our calculation with
$S_T^\sigma = \delta_{\sigma 1}$.
Equation (\ref{eq1}) then reads:
\be
T(x,s_T) = \int \frac{dy_1^-}{4 \pi} e^{i x M y_1^- / \sqrt{2}}
           \bra{p,s_T} \bar{\psi}(0) \gamma^+
           \int dy_2^- \tilde{G}^{1 +} (y_2^-)
           \psi (y_1^-) \ket{p,s_T}  \; .
\ee
Now we make our only approximation by assuming that the integral in
brackets is given by the value of $\tilde{G}^{1 +} (0)$ times the
region of integration, that is:
\be \label{eq2}
\int dy_2^- \; \bar{\psi} \gamma^+ \tilde{G}^{1 +} (y_2^-) \psi
= \bar{\psi} \gamma^+ \tilde{G}^{1 +} (0) \psi \int dy_2^-
= 2 \sqrt{2} c R_0  \bar{\psi} \tilde{G}^{1 +} (0) \psi \; ,
\ee
with the radius of the proton $R_0$.
Depending on how
lightcone integration is defined with respect to the transverse variables
$c$ is a number between $1/3$ and 1.
Figure 1 is meant to visualize a possible configuration for the
chromomagnetic field $\vec{B}$, assuming the quarks to have positive
chromomagnetic moments.
The quark density inside the proton is rather uniform and this should
also be true for the quarks carrying the polarization (in simple quark
models these are in s states).
Then the $\vec{B}$ field should be smooth over the proton volume.
With this picture in mind, equation (\ref{eq2}) seems to be
justified up to a numerical factor of order 1 which we keep in mind
but never write explicitly.
Instead we absorb this factor into the uncertancy of chosing $c$ and $R_0$.
Integration over $T(x,s_T)$ and inserting equation (\ref{eq2})
gives:
\be  \label{eq3}
\int_{-1}^{1} dx  \; T(x,s_T)
= 2 c \frac{R_0}{M} \bra{p,s_T} \bar{\psi}(0) \gamma^+
           \tilde{G}^{1 +} (0) \psi (0) \ket{p,s_T}  \; .
\ee
Here $M$ is the mass of the proton.
Let us now turn to DIS.
Neglecting $Q^2$ dependence
the operator product expansion (OPE) for $g_2$ \cite{RJa91} leads to the
following result.
\be
\int_0^1 x^n g_2 (x) \; dx = - \frac{n}{2 (n + 1)}
\left( a^{(n)} - d^{(n)} \right) \; , \quad n = 2,4, \dots \quad .
\ee
Where the $a^{(n)}$ are known from $g_1$ which will be measured with
high statistics.
\be
a^{(n)} = 2 \int_0^1 x^n g_1 (x) \; dx  \; , \quad n =
0,2,4, \dots \quad .
\ee
Furthermore OPE specifies the correlator which determines $d^{(2)}$.
\be \label{eqneu1}
2 \, d^{(2)} S^{[\sigma} P^{\{\mu_1 ]} P^{\mu_2 \} } - \enspace ({\rm traces})
= - \frac{1}{3} \bra{p,s_T} \bar{\psi} (0) \tilde{G}^{\sigma \{ \mu_1}
  \gamma^{\mu_2 \} } \psi \ket{p,s_T} -  \enspace ({\rm traces})
\ee
Here $ \{ \enspace \} $ means symmetrization of the indices in brackets
$[ \enspace ]$ antisymmetrization of the indices in brackets.
Equation (\ref{eqneu1}) implies
\be
\int_{- 1}^1 T(x, s_T) \; dx =  - 12 c M^2 R_0 \int_0^1 x^2
g_{2; \enspace {\rm tw3 }} (x) \; dx \; .
\ee
This is the basic relation between single photon spin asymmetry and
the polarized structure function $g_2$.
The theoretical predictions for $d^{(2)}$ are rather uncertain. The
only published estimate is that of Balitsky, Braun and Koleschnichenko
\cite{IIBa90}, based on sum rule techniques. They give
\bea \label{eq4}
\int_0^1 dx \; x^2 g_{2; \enspace {\rm tw3 }}^P (x)
&=& - (5 \pm 10 ) \cdot 10^{-4}
\; ,
\nn \\
\int_0^1 dx \; x^2 g_{2; \enspace {\rm tw3 }}^N (x)
&=& - (9 \pm 4 ) \cdot 10^{-3}
\; .
\eea
The superscripts P and N refer to proton and neutron.
Chosing $R_0 = 1 $ fm and $M = 938$ MeV this implies
\bea
\int_{-1}^1 dx \; T^P (x, s_T)
&=&  c (2.7 \pm 5.4 ) \cdot 10^{-2}
\; {\rm GeV} \; ,
\nn \\
\int_{-1}^1 dx \; T^N (x, s_T)
&=&  c (4.6 \pm 2.0 ) \cdot 10^{-1}
\; {\rm GeV} \; .
\eea
We conclude, that a shape for $T (x, s_T)$ like
\be
T (x, s_T) = const \cdot F_2 (x) / x \; {\rm GeV}  \; ,
\ee
as assumed in \cite{JQi91}
is wrong as it leads to an infinite zero'th moment of $T(x,s_T)$.
With the other shape proposed by QS for $T(x,s_T)$, namely
\be
T (x, s_T) = const \cdot F_2 (x) \; {\rm GeV}
\ee
we get
\be
T^P (x, s_T) = (0.08 \pm 0.16) \cdot c \cdot F^P_2 (x) \; {\rm GeV}  \; .
\ee
For the proton this results to positive
asymmetry which is at least a factor of two smaller than
the one proposed by QS.
Note that negative asymmetries are observed for the production of
$\pi^0$ in the same reaction by J. Antille et al. \cite{GBu76}.
A similar conclusion was also reached in \cite{ASc93} based on
completely different arguments.
Relying on the analysis made in
\cite{ASc93}, this result also means, that about 6 \% $ \pm $ 12 \%
of the gluons in the proton are coherently correlated to the spin of the
proton.

However the situation is different for the neutron, where
the integral over $T (x, s_T)$ could be much larger. This could result in
much larger asymmetries for the neutron than those for the proton.
This would be a rather spectacular effect and should be tested
experimentally. Note, however, that the values given in Eq. (\ref{eq4})
are still more ore less compatible with zero.
This stresses the great need to confirm this sum rule results by independent
means.

One can also estimate the matrix element (\ref{eq4}) using bag model
calculations.
These estimations depend very much on the bag renormalization scale.
For the very small value $\mu_{\rm bag} =
250$ MeV the bag model calculations of Stratman \cite{MSt93} fit the
experimental known (polarized) structure functions very well.
The bag model prediction for $d^{(2)}$ differs however strongly from
the sumrule result.
The sign is opposite and for $\mu_{\rm bag} = 866$ MeV,
as used by Ji and Unrau \cite{XJi93}
the bag model values are a factor of nine larger.
Asymmetries resulting from such a high value of the twist three matix
elements (\ref{eq4}) are clearly unphysical, as shown in \cite{ASc93}.

This work was supported in part by DFG (G. Hess Program).
\enspace A.S. thanks also the MPI f\"ur Kernphysik in Heidelberg for its
hospitallity.
L.M. was supported in part by A. v. Humboldt foundation and in part
by KBN under grant 2-0224091-01.
The authors thank G. Sterman for helpful conversation.

\newpage

{\LARGE \bf Figure Caption}
\rm
\\ \\ \\
Figure 1:
Illustration of the possible chromomagnetic field configuration in the proton.
The field is parallel to the transverse spin direction $\vec{S}$ and
the proton moves in the direction $\vec{z}$.
\end{document}